\documentclass[aps,prl,preprint,showpacs,superscriptaddress]{revtex4-1}

\usepackage{amsmath}
\usepackage{epsfig}
\usepackage{epstopdf}
\usepackage{textcomp}
\usepackage{color}

\begin{document}
\title{Point defect states emergence in a plasmonic crystal}

\author{Hikaru Saito}
\email{saito.hikaru.961@m.kyushu-u.ac.jp}
\affiliation{Department of Advanced Materials Science and Engineering, Kyushu University, 6-1 Kasugakoen, Kasuga, Fukuoka 816-8580, Japan}

\author{Hugo Louren\c{c}o-Martins}
\affiliation{Laboratoire de Physique des Solides, Universit\'e Paris-Sud, CNRS-UMR 8502, Orsay 91405, France}

\author{No\'emie Bonnet}
\affiliation{Laboratoire de Physique des Solides, Universit\'e Paris-Sud, CNRS-UMR 8502, Orsay 91405, France}

\author{Xiaoyan Li}
\affiliation{Laboratoire de Physique des Solides, Universit\'e Paris-Sud, CNRS-UMR 8502, Orsay 91405, France}

\author{Tracy C. Lovejoy}
\affiliation{Nion Company, 1102 Eighth Street, Kirkland, Washington 98033, USA}

\author{Niklas Dellby}
\affiliation{Nion Company, 1102 Eighth Street, Kirkland, Washington 98033, USA}

\author{Odile St\'ephan}
\affiliation{Laboratoire de Physique des Solides, Universit\'e Paris-Sud, CNRS-UMR 8502, Orsay 91405, France}

\author{Mathieu Kociak}
\affiliation{Laboratoire de Physique des Solides, Universit\'e Paris-Sud, CNRS-UMR 8502, Orsay 91405, France}

\author{Luiz H. G. Tizei}
\affiliation{Laboratoire de Physique des Solides, Universit\'e Paris-Sud, CNRS-UMR 8502, Orsay 91405, France}

\date{\today}

\begin{abstract}

Plasmonic crystals are well known to have band structure including a bandgap, enabling the control of surface plasmon propagation and confinement. The band dispersion relation of bulk crystals has been generally measured by momentum-resolved spectroscopy using far field optical techniques while the defects introduced in the crystals have separately been investigated by near field imaging techniques so far. Particularly, defect related energy levels introduced in the plasmonic band gap have not been observed experimentally. In order to investigate such a localized mode, we performed electron energy-loss spectroscopy (EELS), on a point defect introduced in a plasmonic crystal made up of flat cylinders protruding out of a metal film and arranged on a triangular lattice. The energy level of the defect mode was observed to lie within the full band-gap energy range. This was confirmed by a momentum-resolved EELS measurement of the band gap performed on the same plasmonic crystal. Furthermore, we experimentally and theoretically investigated the emergence of the defect states by starting with a corral of flat cylinders protrusions and adding sequentially additional shells of those in order to eventually forming a plasmonic band-gap crystal encompassing a single point defect. It is demonstrated that a defect-like state already forms with a crystal made up of only two shells.


\end{abstract}

\maketitle

In periodic metal surface structures, surface plasmon polaritons (SPPs) behave as Bloch waves. In a metal nanodisk array with a triangular lattice on a metal substrate, their energy levels follow an array of cone-shaped dispersion surfaces whose tops are at the reciprocal lattice points under the empty lattice approximation (ELA). Beyond the ELA, mode superposition on the intersections of the dispersion surfaces leads to band-gap formation \cite{Kitson1996, Barnes1996}, in close analogy with electronic band-gap formation in crystals. Following this analogy, several types of defects can be introduced in such a plasmonic band-gap material (PlBG). The SPP dispersion relations of PlBG can be experimentally studied by far-field techniques, provided that the SPP are radiative \cite{VanBeijnum2013,Lozano2013,Hakala2017,Saito2015}. Same applies to certain sorts of defects where energy levels of the defect modes lie in the band gap opening above the light dispersion surface \cite{Saito2015b}. Therefore, the related plasmonic Bloch and defects modes, as well as photonic modes in a photonic band-gap material (PhBG)\cite{Sapienza2012}, can be directly studied by techniques measuring far-field scattering, such as angle resolved angular scattering \cite{VanBeijnum2013,Lozano2013,Hakala2017} or cathodoluminescence \cite{Saito2015,Saito2015b,Sapienza2012}. However, the far-field techniques require dispersion relations lie above the light dispersion surface, hence, their applicability is limited above the second band. Indeed, the first band and the first band gap are located below the light dispersion surface. Additionally, defect modes are expected to be non-radiative if their energy levels lie within the first band gap, and some defects such as point defects are, by analogy with semi-conducting material band structure, very localized in space. The experimental study of such modes is therefore complicated. Although different near field approaches can be used for either measuring the non-radiative part of the band structure of PlBG\cite{Kitson1996} or the defects energy levels \cite{Smolyaninov1999,Bozhevolnyi2002,Marquart2005}, it is at the moment impossible to analyze both at the same time, limiting the understanding we have of these structures and their point defects relation. 
  
It is known from the physics of PhBGs that well-engineered point defects can lead to very high Q states \cite{Akahane2003}. Generating such states in PlBG would allow to increase the magnitude of the electromagnetic energy concentration. Otherwise speaking, one could perform precise engineering of the electromagnetic local density of states (EMLDOS), the key quantity in nanophotonics applications. Indeed, high Q, strongly localized resonances are key to achieve much higher coupling to localized quantum emitters such as quantum dots \cite{Santhosh2016}, molecules \cite{Wertz2015} with potential applications in sensing for example. Therefore, the characterization and understanding of the formation of point defect inside well-characterized PlBG is required.
   
We hereby report on the investigation of the formation of energy levels by point defects in a plasmonic crystal using electron energy loss spectroscopy (EELS). EELS signal is closely related to the electromagnetic density of states and is well-known to be sensitive to both radiative and non-radiative excitations \cite{Losquin2015,Losquin2015b}. EELS can access all transverse magnetic modes over the entire reciprocal space, as it has been shown more than 40 years ago by Silcox and coworkers \cite{Pettit1975,Chen1975}. For SPP Bloch waves, the momentum transfer reaches the photonic range, i.e., the required momentum resolution is thousand times higher than that for electrons in the solid-state materials. Such a high momentum resolution has been achieved in the non-monochromated electron microscopes so far \cite{Pettit1975, Chen1975,Midgley1999,Saito2015c}. With that said, the band dispersion relation of photonic band gap materials or plasmonic band gap materials has not yet been experimentally investigated by EELS, although predicted for photonic modes in pioneering works \cite{Garcia2003,Garcia2003b}. Here, we used a newly-developed monochromated electron microscope \cite{Krivanek2014} with a high electron optics stability. In this electron microscope, highly spatially-resolved and momentum-resolved spectroscopy modes are switchable by changing the electron optics configuration. Therefore, point defect modes were investigated with a nanometer spatial resolution, and the band dispersion relation of the surrounding bulk crystal was measured in the same sample with a order of 0.001 nm$^{-1}$ momentum resolution. Furthermore, the emergence of the defect states was experimentally investigated by building a crystal layer by layer around a vacancy.   
   
The PlBGs with a triangular lattice with a 390 nm period were used in the experiments. Using e-beam lithography, we produced on a 15 nm thin Si$_3$N$_4$ film an array of aluminum disks with a 100 nm height (confirmed by EELS measurements \cite{Egerton1996}) and a diameter of 260 nm arranged with a 390 nm period. Then, another metal layer was deposited on the top of the disk array to make a continuous corrugated metal surface where a plasmonic full bandgap has been demonstrated in the past \cite{Kitson1996}. A full bandgap opens in this kind of surface structures when the energy value of the lower band edge at the K point becomes lower than that of the higher band edge at the $M$ point. The basic band structure and the band edge modes at the $M$ and $K$ points are described under the empty lattice approximation (ELA) in Supplemental Material. 
   
\begin{figure}
\includegraphics[width=1\columnwidth]{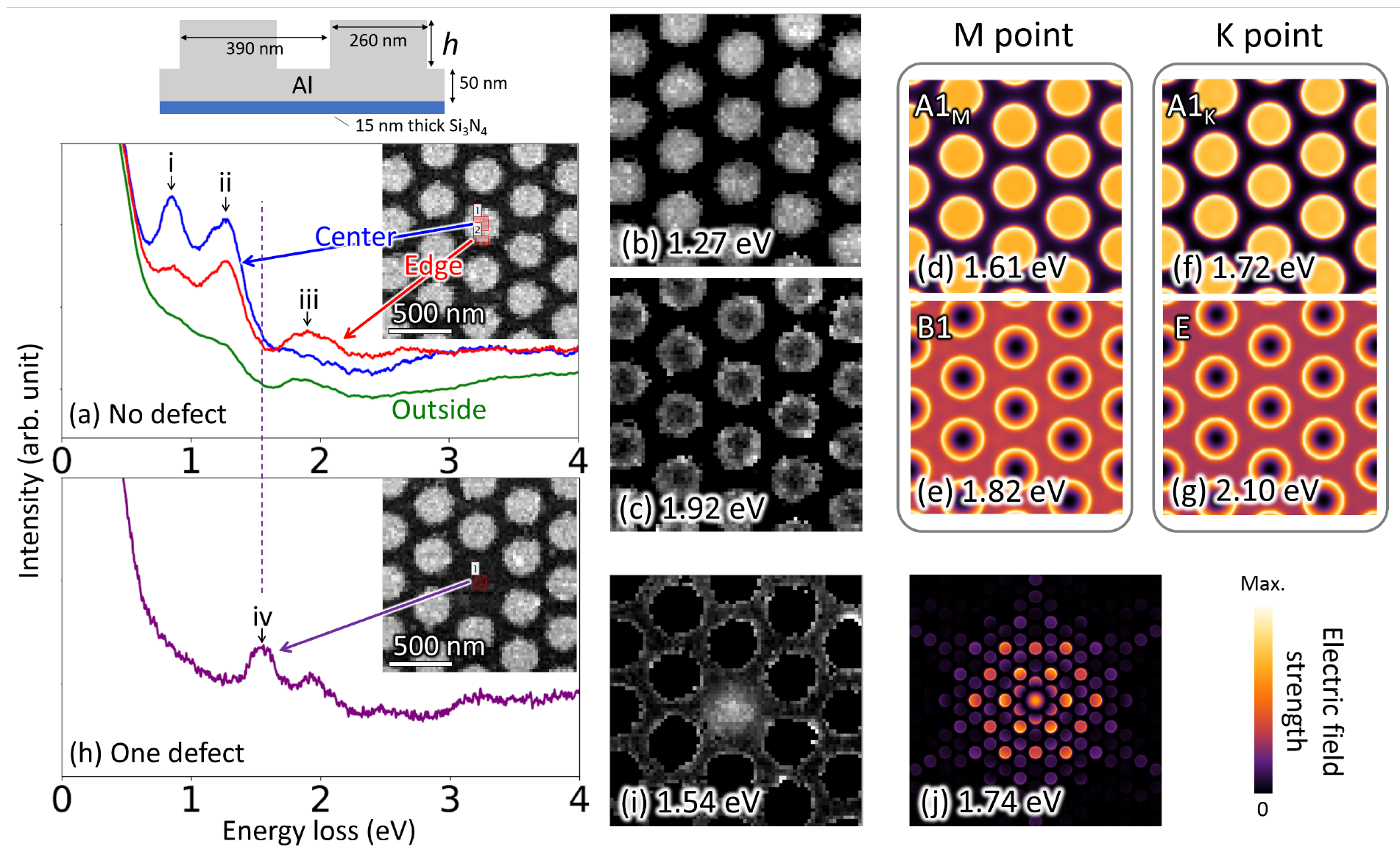}
\caption{\label{figure1} (a) Spatially-resolved EELS spectra extracted from the centers (blue), edges (red) and outside (green) of disks on the PlBG. The blue and red spectra were averaged by using plural spectra extracted from the equivalent sites. One of the extraction positions is indicated in the inset annular dark field image. The green spectrum was averaged over the whole parts of the outside of the disks shown in the inset. (b,c) Energy-filtered images at (b) 1.27 eV and (c) 1.92 eV for the plasmonic crystal. (d-g) Surface normal component of electric field strength at the band edges of the M and K points calculated by FDTD method for h = 60 nm, which which are averaged based on the translational and the rotational symmetries. (h,i) EELS spectrum extracted from the point defect shown in the inset and corresponding energy-filtered images. (j) Surface normal component of electric field strength of the defect mode calculated by FDTD method.}

\end{figure}

We first performed spatially-resolved EELS experiment  using a monochromated scanning transmission electron microscope (STEM) Nion Hermes 200-S. It was operated at an acceleration voltage of 200 kV with a subnanometer spatial resolution and an energy resolution of 70 meV. Fig. \ref{figure1}a-c show the experimental results taken from a bulk crystal whose top surface is covered with aluminum. Three peaks (numbered i, ii, iii) are seen in the extracted spectra, whose relative intensities change with beam position as shown in Fig. \ref{figure1}a. Among them, the peak i is not a Bloch mode but a radial breathing mode localized at each single disk \cite{Schmidt2012} as discussed in Supplemental Material. The peaks ii and iii come from the local maxima of the plasmonic density of states at the lower and upper band edges. As shown in the energy-filtered images (Fig. \ref{figure1}b and c), the excitation probabilities of the lower band edge modes are distributed to the entire top surface of the disks while those of the upper band edge modes are localized to the edges of the disks, indicating the same results as the previous study, that is, the lattice point modes are located at the lower band edges while the inter-lattice modes are located at the upper band edges \cite{Yoshimoto2018}. Figs. \ref{figure1}d-g show simulation results for the band edge modes at the $M$ and $K$ points performed by finite difference time domain (FDTD) calculations, indicating the same trend as the experimental results, i.e., both lattice point modes A1$_M$ and A1$_K$ exhibit energy levels lower than the inter-lattice point modes B1 and E. However, the band edges at the $M$ and $K$ points does not seem to be resolved by real space mapping where the momentum transfer is integrated. Their spatial distributions are also similar to each other as suggested in FDTD simulations, meaning that momentum-resolved spectroscopy is the only way to distinguish them. 

Fig. \ref{figure1}h shows an EELS spectra taken from a point defect deliberately introduced into the PlBG. The observed new peak iv is located between the peaks ii and iii (Fig. \ref{figure1}a). In energy-filtered image corresponding to the new peak (Fig. \ref{figure1}i), a localized intensity distribution is seen near the point defect, suggesting that this PlBG prevents SPPs from propagating in the crystal in the bandgap energy range. Otherwise speaking, this point defect behaves as a cavity, an analogue of defects in photonic crystals \cite{Sapienza2012, Akahane2003}. Since the antinode of the observed defect mode is at the original lattice point position, it is inferred that it is formed due to modification of the lattice point mode of the bulk crystal. This result is reproduced in the FDTD simulation (Fig. \ref{figure1}j) and indicating that the electric field distribution is similar to A1$_K$ mode (Supplemental Material), leading to a consideration of the following formation mechanism of the defect mode. Since SPP plane waves on the flat surface follow the dispersion relation with a longer wavelength compared to the lattice point mode, the introduction of the point defect locally makes the effective refractive index lower, elevating the energy levels of the lattice point modes. As a result, the energy level of the A1$_K$ mode, which lies at the highest energy level of the first band, enters within the full band gap, and thus, the modified A1$_K$ mode is localized near the point defect.

\begin{figure}
\includegraphics[width=0.6\columnwidth]{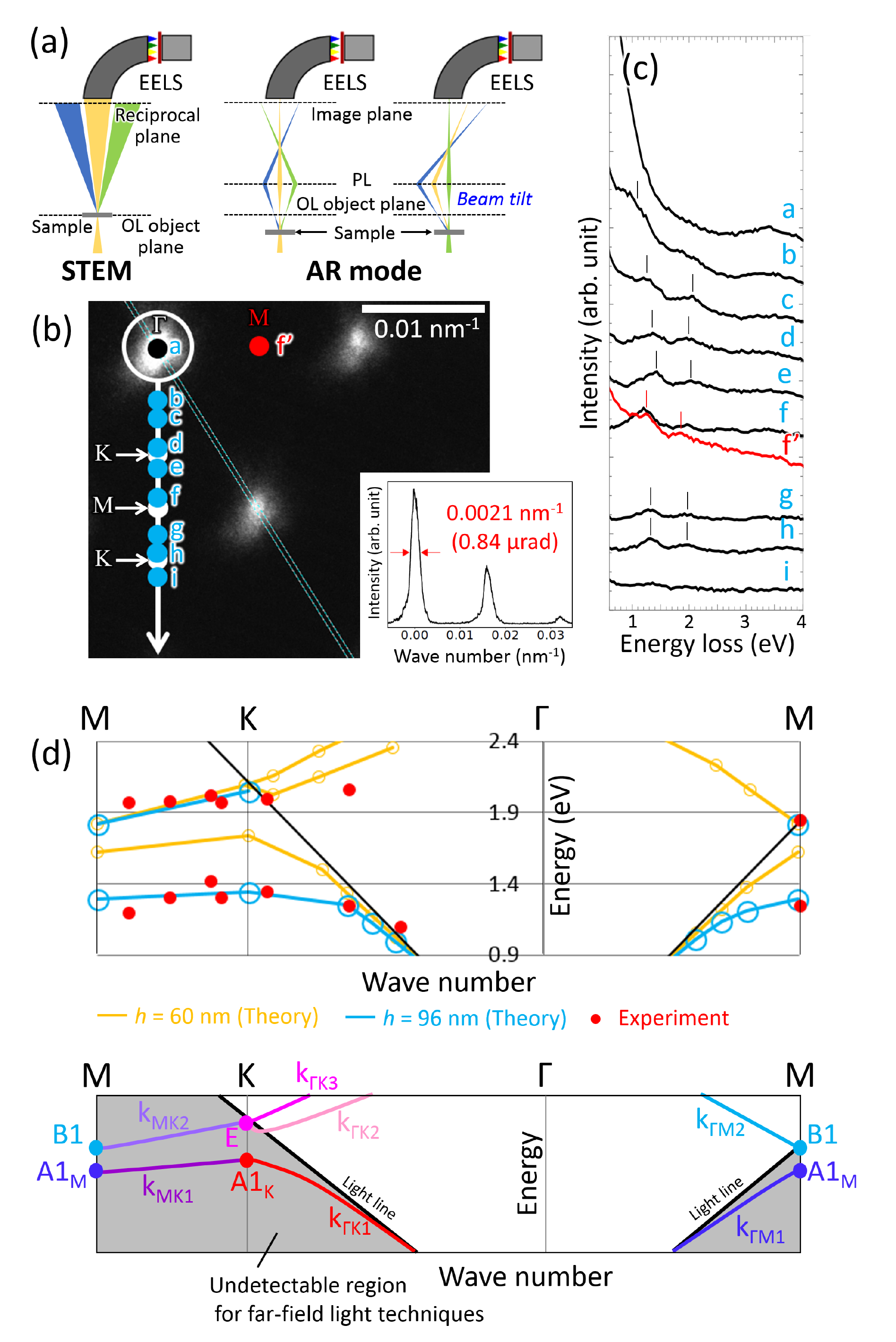}
\caption{\label{figure2} (a) Experimental configurations for STEM-EELS (left) and the angle resolved (AR) mode (right). (b) Diffraction pattern taken from the sample in the AR mode (left) and a profile extracted along the line depicted in the left pattern (right). (c) Series of momentum-resolved EELS spectra taken by shifting the diffraction pattern relative to the entrance aperture of the EELS spectrometer. The sampling position of each spectrum is indicated in Fig. \ref{figure1}b. The diameter of the entrance aperture is depicted as the white circle (Fig. \ref{figure1}b). (d) Comparison of the dispersion relations between the momentum-resolved EELS experiment (red circles) and the FDTD simulations for h = 60 nm (yellow circles) and 96 nm (aqua circles). The black line is light dispersion. The schematic drawing for each band dispersion curve is at the bottom. }

\end{figure}

\begin{figure}
\includegraphics[width=1\columnwidth]{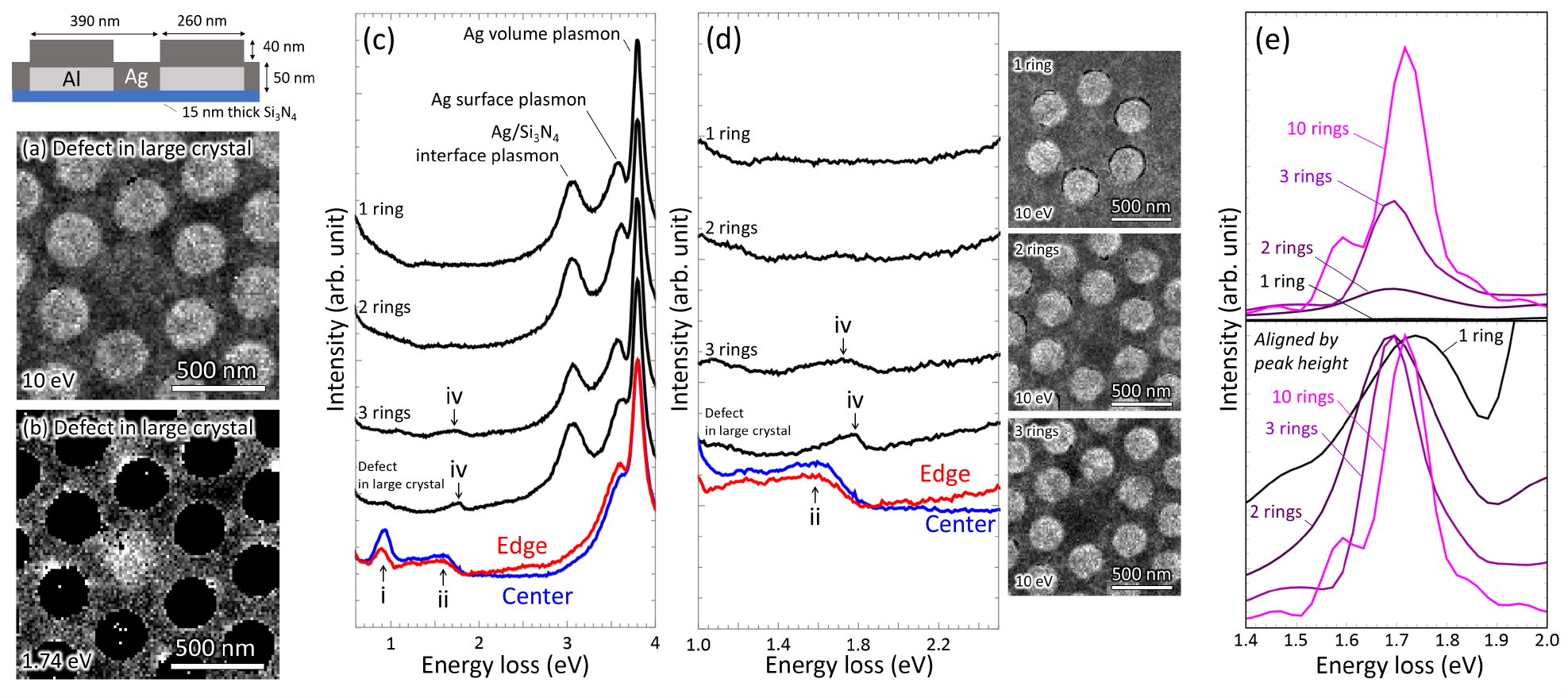}
\caption{\label{figure3} (a) Energy-filtered image of 10 eV taken from a point defect surrounded by a large crystal. (b) Energy-filtered image of the defect mode obtained from the sample corresponding to (a). (c) EELS spectra extracted from the point defect surrounded by different sizes of crystals. The blue and red spectra are extracted from centers and edges of the disks in a large crystal with no defect, which are shown for comparison. (d) Magnified EELS spectra of (c) (left). Corresponding Energy-filtered images of 10 eV for the 1, 2 and 3 rings are shown in the right. (e) Spectra of the defect modes simulated by FDTD method. The detailed parameters used for the simulations are written in Supplemental Material.}

\end{figure}

In order to confirm our assumption over the energy position of the band gap, we performed  momentum-resolved EELS on the same sample by changing the electron optics configuration to the so-called “angular-resolved (AR) mode” as shown in Fig. \ref{figure2}a. To achieve a very high momentum resolution and a moderate spatial resolution simultaneously, the configuration proposed by Midgley \cite{Midgley1999} was adopted in this experiment. As shown in Fig. \ref{figure3}a, the specimen height was moved away from the standard height, leading to the formation of a diffraction pattern at the object plane. Post-specimen projector lenses were used to couple this diffraction pattern, magnified, to the plane at the entrance of the EELS spectrometer. A scattering vector range was selected by displacing the entrance aperture of the spectrometer on the magnified diffraction plane. Fig. \ref{figure2}b shows a diffraction pattern obtained from the bulk crystal with the period of 390 nm. Thanks to the large pole piece gap of 6 mm, a large specimen height change of 1 mm from the standard height was available, resulting in a very high momentum resolution 0.0021 nm$^{-1}$. Considering the convergence semi-angle of 5 mrad, the illuminated area on the sample was 10 $\mu$m in diameter.
   
Momentum-resolved spectra obtained at an acceleration voltage of 200 kV and an energy resolution of 100 meV are shown in Fig. \ref{figure2}c. The points on the diffraction pattern in Fig. \ref{figure2}b indicate the centers of the entrance aperture when acquiring the corresponding EELS spectra. The diameter of the entrance aperture was 0.0056 nm$^{-1}$ in the diffraction plane, corresponding to the white circle in the Fig. \ref{figure2}b. Although no peak is observed below 3 eV in the spectrum acquired on the $\Gamma$ point, as the measurement point approaches the $K$ point, two peaks appear. The first peak is located below the energy value for $K$ point under the ELA (2.04 eV), meaning that this peak is attributed to the lowest energy band. The second peak is located around 2 eV or more near the $K$ point. When the measurement point passes the $K$ point and comes closer to the point $M$, the peak shifts to the lower energy side as shown in the spectrum f, indicating that the energy levels of band edge at the M point are lower than those at the $K$ point. This behavior agrees with the FDTD simulations (Figs. \ref{figure1}d-g) and is also observed in the spectrum f' taken from another $M$ point. When the measurement point passes through the point $M$ and reaches the point $K$ again, the peak shifts again to the high energy side as shown in the spectra g and h. Fig. \ref{figure2}d compares the dispersion relations obtained from the experiment and FDTD simulations. Unlike momentum-resolved CL spectroscopy in the past \cite{Saito2017}, nonradiative modes below the light line are also detected in this momentum-resolved EELS. Only one branch is detected in CL experiment although there are two branches $\boldsymbol{k}_{\Gamma K2}$ and $\boldsymbol{k}_{\Gamma K3}$. The dispersion curve of the $\boldsymbol{k}_{\Gamma K2}$ branch is flatter than that of the $\boldsymbol{k}_{\Gamma K3}$ branch, making its contribution larger in the spectra near the K point. Accordingly, the lower and upper band edges have been detected by momentum-resolved EELS, which appeared at the similar energy positions as the peaks ii and iii detected in momentum-integrated EELS (Fig. \ref{figure1}a). It has been confirmed that the observed peak iv at the point defect (Fig. \ref{figure1}h) was located within the full bandgap of the bulk crystal.

Finally, the amplitude of the defect mode depends on the size of the PlBG (Fig. \ref{figure3}) . This gives an experimental proof of SPP confinement due to the bandgap.  The size dependence is evidenced by experiments with point defects surrounded by different numbers of disk arrays rings. We call a ring each group of disks that completely surround the perimeter of the defect maintaining the crystal symmetry. Here, the cover metal is changed to silver in anticipation that the ohmic loss of the SPP is reduced \cite{Ross2014}. The disk height was adjusted to be lower than the sample of Figs. \ref{figure1} and \ref{figure2} in consideration of the decrease in electron beam penetration associated with the use of heavy metal. In the case of large crystal where the 10 unit cells surrounds the point defect, an antinode is observed at the defect in the energy-filtered image (Fig. \ref{figure3}b), reproducing the result shown in Fig. \ref{figure1}i. When the crystal size surrounding the defect decreases to 3 rings (Figs. \ref{figure3}c-d), the peak gets broad although the peak of the defect mode can still be confirmed, indicating increase of leakage to the outside of the crystal.
   
The peak of defect mode is hardly recognized when the crystal size gets smaller than 2 rings, indicating drastic reduction of the band gap effect depending on the ring number. This peak broadening and reduction of peak intensity accompanied with crystal size down are well reproduced by FDTD simulations shown in Fig. \ref{figure3}e. It should be noted that the peak position gradually shifts to higher energy side in the cases of 2 or more ring numbers but that of the 1 ring is located at higher energy side than that of the 10 rings. This fact means that modal formation mechanism suddenly changes with emergence of the periodicity of the ring structure.

In summary, the relationship between the band structure of the PlBG and the defect mode was revealed by combining spatially-resolved spectroscopy and momentum resolved spectroscopy of EELS, giving the formation model of the defect mode, i.e., the energy level of the band-edge mode lying at the highest level of the first band is elevated by introduction of the flat area, resulting in formation of the modified mode within the band gap energy range. Furthermore, it has been confirmed that the excitation amplitude and quality factor of the defect mode depends on the crystal size. However, the defect mode is already formed in the 2 ring structure, which coincides with emergence of periodicity of the ring structure.

\textbf{Conflicts of Interest} N.D., and T.C.L. have a financial interest in Nion Co. The other authors declare no conflict of interest.

\begin{acknowledgments}
This work was supported by Japan Society for the Promotion of Science Kakenhi No. 17K14118, The Murata Science Foundation and the program of future investment TEMPOS-CHROMATEM (No. ANR-10-EQPX-50).
\end{acknowledgments}

\end{document}